\documentclass[11pt]{article}
\usepackage{moriond,epsfig}

\usepackage{axodraw}

\begin{document}
\vspace*{4cm}
\title{PREDICTION OF PURE ANNIHILATION TYPE B DECAYS}

\author{C.D. L\"u }

\address{Institute of High Energy Physics, CAS, P.O.Box 918(4),
Beijing 100039, China
}
\maketitle

\abstracts{
  The rare decays $B^+ \to D_s^{+} \phi$, $B^0  \to D_s^{(*)-} K^+$
 and $B^+ \to D_s^{(*)+} \overline{K}^0$ can occur only via annihilation type
 diagrams in the standard model. We calculate these decays in perturbative
 QCD approach. We found that the calculated branching ratio of $B^0 \to D_s^-
 K^+$
 agree with the data which had been observed in the KEK and SLAC B factories.
 The decay $B^+ \to D_s^{(*)+} \bar K^0$ has a very small branching ratio at
 O ($10^{-8}$), due to the suppression from CKM matrix elements.
 The branching ratio of $B^+ \to D_s^+ \phi$ is of order $10^{-7}$ which may
 be measured in the near future by KEK and SLAC B factories. The small
 branching ratios predicted in the standard model make these channel sensitive to
 new physics contributions.}

\section{Introduction}

The generalized factorization approach has been applied to the
theoretical treatment of non-leptonic  $B$ decays for years \cite{bsw}.
It is a great success in explaining many decay branching
ratios\cite{akl1}.
The factorization approach (FA) is a rather simple method.
Some efforts have been made to improve their theoretical
application\cite{bbns} and to understand the reason why the FA
has gone well\cite{Keum:2000}.
One of these methods is the perturbative QCD approach (PQCD),
where we can calculate the annihilation diagrams as well as the factorizable
and nonfactorizable diagrams.

The rare decays $B \to D_s^{(*)} K (\phi)$ are pure annihilation type decays.
In the usual FA, this decay picture is described as
$B$ meson annihilating into vacuum and the $D_s^{(*)}$ and $K(\phi)$ mesons produced
from vacuum then afterwards.
To calculate these decays in the FA, one needs the
$D_s^{(*)} \to K(\phi)$ form factor at very large time like momentum transfer
${\cal O} (M_B)$.
However the form factor at such a large momentum transfer is not known
in FA.
The annihilation amplitude is a phenomenological parameter in QCD
factorization approach (QCDF)\cite{bbns}, and the QCDF calculation of
these decays is also unreliable.
Here, we will try to use the PQCD approach,
to evaluate the $B \to D_s K (\phi)$ decays.
By comparing the predictions with the experimental data,
we can test the PQCD evaluation of the annihilation amplitude.

A $W$ boson exchange causes $\bar{b}d \to \bar{c}u $ or
$\bar b u \to \bar{d} (\bar s )c$, which is usually described by the effective four quark
operators, and
the additional $\bar{s}s$ quarks included in $D_s K (\phi) $ are produced from a gluon.
This gluon  attaches to any one of the quarks participating in the
four quark operator.
In the rest frame of $B$ meson,
 $s$ and $\bar{s}$ quarks included in $D_s K(\phi)$ each has
$\mathcal{O}(M_B/4)$ momenta, and the gluon producing them   has
$q^2 = \mathcal{O}(M_B^2/4)$. This is a hard gluon.
One can perturbatively treat the process where the four quark operator
 exchanges a hard gluon with $s \bar s$ quark pair.
Therefore the quark picture becomes six-quark interactions. The
decay amplitude is then expressed as product of the hard six
quark operators and the non-perturbative meson wave functions.

\section{Framework}\label{sc:fm}

 PQCD approach has been developed and applied in the non-leptonic $B$
meson decays\cite{Chang:1997dw,Keum:2000,PQCD}
for some time.
In this approach,
the decay amplitude is separated into
soft($\Phi$), hard($H$), and harder($C$) dynamics characterized by
different scales.
It is conceptually written as the convolution,
\begin{equation}
A \sim \int\!\! d^4k_1 d^4k_2 d^4k_3\
\mathrm{Tr} \bigl[ C(t) \Phi_B(k_1) \Phi_{D_s}(k_2) \Phi_K(k_3)
H(k_1,k_2,k_3, t) e^{-S(t)}\bigr],
\label{eq:convolution1}
\end{equation}
where $k_i$'s are momenta of light quarks included in each mesons, and
$\mathrm{Tr}$ denotes the trace over Dirac and color indices.
$C(t)$ is Wilson coefficient of the four quark operator.
In the above convolution, $C(t)$ includes the harder dynamics
at larger scale than $M_B$ scale and describes the evolution of local
$4$-Fermi operators from $m_W$,   down to the scale $t$,
where $t= \mathcal{O}(\sqrt{\bar{\Lambda} M_B})$.
$H$ describes the four quark operator and the spectator quark connected by
 a hard gluon whose $q^2$ is at the order
of $t$, and includes the hard dynamics characterized by the scale $t$.
Therefore, this hard part $H$ can be perturbatively calculated, which is process dependent.
$\Phi_M$ is the wave function which describes hadronization of the quark
and anti-quark to the meson $M$.
$\Phi_M$ is  independent of the specific processes.
Determining $\Phi_M$ in some other decays,
we can make quantitative predictions here.

The large double logarithms ($\ln^2 x_i$) on the longitudinal direction
are summed by the threshold resummation\cite{L3},
and they lead to $S_t(x_i)$ which smears the end-point singularities
on $x_i$.
The last term, $e^{-S(t)}$, contains two kinds of logarithms.
One of the large logarithms is due to the renormalization of
ultra-violet divergence $\ln tb$, the other is double logarithm
$\ln^2 b$ from the overlap of collinear and soft gluon corrections.
This Sudakov form factor suppresses the soft dynamics effectively\cite{soft}.
Thus it makes perturbative calculation of the hard part $H$ applicable
at intermediate scale, i.e., $M_B$ scale.

In general, $\Phi_{M,\alpha\beta}$ having Dirac indices $\alpha,\beta$
are decomposed into 16 independent components, $1_{\alpha\beta}$,
$\gamma^\mu_{\alpha\beta}$, $(\gamma_5\sigma^{\mu\nu})_{\alpha\beta}$,
$(\gamma^\mu\gamma_5)_{\alpha\beta}$, $\gamma_{5\alpha\beta}$.
If the considered meson $M$ is $B$ or $D_s^{(*)}$ meson,
to be pseudo-scalar and heavy meson, the structure
$(\gamma^\mu\gamma_5)_{\alpha\beta}$ and $\gamma_{5\alpha\beta}$
components remain as leading contributions.
Then, $\Phi_{M,\alpha\beta}$ is written by
\begin{equation}
 \Phi_{M,\alpha\beta} = \frac{i}{\sqrt{6}}
\left\{
(\not \! P_M \gamma_5)_{\alpha\beta} \phi_M^A
+ \gamma_{5\alpha\beta} \phi_M^P
\right\}.
\end{equation}
As heavy quark effective theory leads to
$\phi_B^P \simeq M_B \phi_B^A$, we have only one independent
distribution amplitude for B meson.
 The heavy $D_s$ meson's wave function can also  be derived
 similarly.

         \begin{figure}[htbp]
         %\begin{minipage}[c]{2.5cm}
      %--------------------------------------------------------
      %  O6 factorizable annihilation diagrams (Mg)
      %--------------------------------------------------------
       \scalebox{0.7}{
         \begin{picture}(130,110)(0,-10)
            \ArrowLine(50,50)(13,63)
            \ArrowLine(13,37)(50,50)
            \Line(55,50)(72.5,32.5)
            \ArrowLine(72.5,32.5)(95,10)
            \Line(55,50)(72.5,67.5)
            \ArrowLine(95,90)(72.5,67.5)
            \ArrowLine(90,50)(115,75)
            \ArrowLine(115,25)(90,50)
            \Gluon(72.5,67.5)(90,50){3}{4}
            \Vertex(90,50){1.5} \Vertex(72.5,67.5){1.5}
            \put(0,47){$B^+$}
            \put(108,7){$D_s^+$}
            \put(108,89){$\phi$}
                \put(20,70){\small{$\bar{b}$}}
                     \put(18,28){\small{$u$}}
                  \put(75,14){\small{$c$}}
                         \put(77,84){\small{$\bar s$}}
            \put(50,-5){(a)}
         \end{picture}
       }
   %--------------------------------------------------------
   %  O6 factorizable annihilation diagrams (Mh)
   %--------------------------------------------------------
    \scalebox{0.7}{
      \begin{picture}(130,110)(0,-10)
         \ArrowLine(50,50)(13,63)
         \ArrowLine(13,37)(50,50)
         \Line(55,50)(72.5,32.5)
         \ArrowLine(72.5,32.5)(95,10)
         \Line(55,50)(72.5,67.5)
         \ArrowLine(95,90)(72.5,67.5)
         \ArrowLine(90,50)(115,75)
         \ArrowLine(115,25)(90,50)
         \Gluon(72.5,32.5)(90,50){3}{4}
         \Vertex(90,50){1.5} \Vertex(72.5,32.5){1.5}
         \put(0,47){$B^+$}
         \put(108,7){$D_s^+$}
         \put(108,89){$\phi$}
   %      \put(42,55){\rotatebox{90}{\small{$S+P$}}}
   %      \put(53,55){\rotatebox{90}{\small{$S-P$}}}
         \put(50,-5){(b)}
      \end{picture}
    }
    \scalebox{0.7}{
      \begin{picture}(120,130)(0,-20)
            \ArrowLine(50,50)(13,63)
            \ArrowLine(13,37)(26,41.8)      \ArrowLine(26,41.8)(50,50)
            \Line(55,50)(72.5,32.5)
            \ArrowLine(72.5,32.5)(95,10)
            \Line(55,50)(72.5,67.5)
            \ArrowLine(95,90)(72.5,67.5)
            \ArrowLine(90,50)(115,75)
            \ArrowLine(115,25)(90,50)
                       \GlueArc(46,160)(120,261,292){4}{9}
            \Vertex(90,50){1.5} \Vertex(26,41.8){1.5}
         \put(0,45){$B^+$}
         \put(105,3){$D_s^+$}
         \put(105,87){$\phi$}
   %      \put(60,38){\small{$V-A$}}
   %      \put(60,55){\small{$V-A$}}
         \put(20,70){\small{$\bar{b}$}}
              \put(18,28){\small{$u$}}
           \put(75,14){\small{$c$}}
                   \put(77,84){\small{$\bar s$}}
         \put(50,-20){(c)}
      \end{picture}
    }
   %--------------------------------------------------------
   %  non-factorizable annihilation diagrams (Mf)
   %--------------------------------------------------------
     \scalebox{0.7}{
       \begin{picture}(120,130)(0,-20)
             \ArrowLine(50,50)(26.5,57.9)    \ArrowLine(26.5,57.9)(13,63)
             \ArrowLine(13,37)(50,50)
             \Line(55,50)(72.5,32.5)
             \ArrowLine(72.5,32.5)(95,10)
             \Line(55,50)(72.5,67.5)
             \ArrowLine(95,90)(72.5,67.5)
             \ArrowLine(90,50)(115,75)
             \ArrowLine(115,25)(90,50)
                        \GlueArc(46.2,-60.5)(120,68.5,99){4}{9}
             \Vertex(90,50){1.5} \Vertex(26.5,57.9){1.5}
          \put(0,45){$B^+$}
          \put(105,3){$D_s^+$}
          \put(105,87){$\phi$}
    %      \put(60,38){\small{$V-A$}}
    %      \put(60,55){\small{$V-A$}}
          \put(20,70){\small{$\bar{b}$}}
               \put(18,28){\small{$u$}}
            \put(75,14){\small{$c$}}
                    \put(77,84){\small{$\bar s$}}
          \put(50,-20){(d)}
       \end{picture}
     }
    \caption{Diagrams for $B^+ \to D_s^+ \phi$ decay. The factorizable
    diagrams (a),(b), and non-factorizable (c),
    (d).}
    \label{fig2}
   \end{figure}
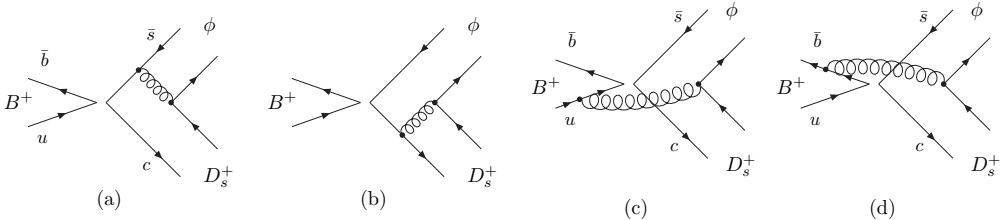

In contrast to the $B$ and $D_s$ mesons,
for the $K$ meson, being light meson,
the $\gamma_5 \sigma^{\mu\nu}$ component remains.
Then, $K$ meson's wave function is parameterized  as
\begin{equation}
 \Phi_{K, ij}(x_3,b_3)
= \frac{i\delta_{ij}}{\sqrt{2N_c}}
\Bigl[ \gamma_5 \not \! P_3 \phi_K^A(x_3,b_3)
+ m_{0K} \gamma_5 \phi_K^P(x_3,b_3)
 + m_{0K} \gamma_5 (\not v \not n - 1)\phi_K^T(x_3,b_3)
\Bigr]\;,
\end{equation}
where $m_{0K} = M_K^2/(m_u + m_s)$, $v = (0,1,{\bf 0}_T ) $,
$n = (1,0,{\bf 0}_T)$.
In $B\to D_s \phi$ decay, only longitudinal polarization of the
$\phi$ meson wave function is relevant, which is similar to K
meson \cite{lu}.

There are four kinds of Feynman diagrams contributing to the six
quark hard dynamics, which is shown in Fig.1. the calculation of
the hard parts are tedious and channel dependent. the results are
shown  in ref.\cite{lu,lu2}.

 \section{Numerical evaluation}\label{sc:neval}

 In this section we show numerical results.
 First for the $B$  meson's wave function,
  we use the same distribution amplitude as
   adopted in Ref.~\cite{Keum:2000}.
 This choice of $B$ meson's wave function is almost a best fit from
 the $B\to K\pi$, $\pi \pi$ decays.
For the $D_s^{(*)}$ meson's wave function,
 we assume the same form as $D^{(*)}$ meson's
 one \cite{h2h}.
 The wave functions $\phi_K^{A,P,T}$ of the $K$ meson are expanded by
 Gegenbauer polynomials, which
 are given in Ref.~\cite{Ball:1998je}.

 For the neutral decay $B^0 \to D_s^{(*)-} K^+$, the dominant contribution is
 the nonfactorizable annihilation diagrams, which
  is proportional
 to the Wilson coefficient $C_2(t)\sim 1$. The factorizable
 annihilation diagram contribution is proportional to $a_2=C_1+C_2/3$,
 which is one order magnitude smaller.
 For the charged decay $B^+ \to D_s^{(*)+}\overline{K}^0 (\phi)$, it is the inverse
 situation.

 The propagators of inner quark and gluon in
 FIG.~\ref{fig2}   are usually
 proportional to  $1/x_i$.
 One may suspect that these amplitudes are enhanced by the endpoint
 singularity around $x_i \sim 0$.
 However this is not the case in
 our calculation. First we introduce the transverse momentum of
 quark, such that the propagators become $1/(x_i x_j +k_T^2)$.
 Secondly, the Sudakov form factor $\mathrm{Exp}[-S]$ suppresses the region
 of small $k^2_T$. Therefore there is no singularity in our
 calculation. The dominant contribution is not from the endpoint of the
 wave function. As a proof,
 in our numerical calculations, for example, an expectation value of $\alpha_s$
 in the integration results in
 $\langle \alpha_s/\pi \rangle = 0.10$,
 Therefore, the perturbative calculations are self-consistent.

 The predicted branching ratios are    \cite{lu}
   \begin{eqnarray}
   \mathrm{Br}(B^0 \to D_s^- K^+)
   = (3.1\pm 1.0) \times 10^{-5}
   , ~~~~~~~~~
   \mathrm{Br}(B^+ \to D_s^+ \overline{K}^0)
   = (1.7\pm 0.4 ) \times 10^{-8}  ,  \\
      \mathrm{Br}(B^0 \to D_s^{*-} K^+)
      = (2.7\pm 0.6) \times 10^{-5}
      ,   ~~~~~~~~
      \mathrm{Br}(B^+ \to D_s^{*+} \overline{K}^0)
      = (4.0\pm 0.8 ) \times 10^{-8}
      ,
   \end{eqnarray}
for variation of the input parameters of wave functions.
 They agree with the experimental observation
 by Belle\cite{Krokovny:2002pe}
 and BaBar\cite{Aubert:2002eu},
 \begin{eqnarray}
  \mathrm{Br}(B^0 \to D_s^- K^+)
 & = &(4.6{}^{+1.2}_{-1.1}\pm 1.3) \times 10^{-5},  ~~~~~~~~ \mbox{Belle, } \\
  \mathrm{Br}(B^0 \to D_s^- K^+)
 & = & (3.2\pm 1.0 \pm 1.0) \times 10^{-5},  ~~~~~ \mbox{BaBar, }
 \end{eqnarray}
 and the experimental  upper limit given
 at $90$\% confidence level\cite{pdg}:
 $
  \mathrm{Br}(B^+ \to D_s^+ \overline{K}^0) < 1.1 \times 10^{-3}$.
  For $B^+ \to D_s^+ \phi$, the predicted branching ratio is
  \cite{lu2}
 $
   \mathrm{Br}(B^+ \to D_s^+ \phi) = 3.0 \times 10^{-7},
  $
  which is still far from  the current experimental upper limit
  \cite{pdg}:
  $ \mathrm{Br}(B^+ \to D_s^+ \phi ) < 3.2 \times 10^{-4}$.

 Despite the calculated perturbative annihilation contributions,
 there is also hadronic picture for the $B^0 \to D_s^- K^+$ decay:
 $B^0 \to D^- \pi^+(\rho^+) \to D_s^- K^+$ through final
 state interaction.  Our numerical results show that the PQCD
 contribution to this decay is already enough to account for the
 experimental measurement. It implies that the soft final state
 interaction is      not important in the  $B^0 \to D_s^- K^+$
 decay. This is consistent with the argument in Ref.~\cite{cl}.
 We expect the same situation happens in other decay channels.

 \section{Conclusion}

 In two-body   $B$   decays, the final state mesons are
 moving very fast, since each of them carry more than 2 GeV energy.
 There is not enough time for them to exchange soft gluons.
 The soft final state interaction may not be important.
  This is consistent with the argument based on
 color-transparency\cite{cl}.
 The PQCD with Sudakov form factor is a self-consistent approach to
 describe the two-body $B$ meson decays.
 Although the  annihilation diagrams are suppressed comparing to
 other spectator diagrams, but their contributions are not
 negligible in PQCD approach\cite{Keum:2000}.

We calculate the $B^0 \to D_s^{(*)-} K^+$ and
 $B^+ \to D_s^{(*)+} \overline{K}^0  (\phi)$ decays, which  occur purely via
 annihilation type diagrams.
 The branching ratio of
  $B^0 \to D_s^- K^+$ decay is sizable, which  has been observed in the $B$
 factories\cite{Krokovny:2002pe,Aubert:2002eu}.
The predicted branching ratio is in good agreement
 with the data.

\section*{Acknowledgments}

We  thank the organizers of the
conference for local support.
This work is partly supported by National Science Foundation of
China under Grant (No. 90103013 and 10135060).


\section*{References}
\begin{thebibliography}{99}

 \bibitem{bsw}M. Wirbel, B. Stech, M. Bauer, Z. Phys. C29, 637 (1985);
  M. Bauer, B. Stech, M. Wirbel, Z. Phys. C34, 103 (1987);
 L.-L. Chau, H.-Y. Cheng, W.K. Sze, H. Yao, B.
 Tseng, Phys. Rev. D43, 2176 (1991), Erratum: D58, 019902 (1998).

 \bibitem{akl1} A. Ali, G. Kramer and C.D. L\"u, Phys. Rev. D58, 094009
 (1998); C.D. L\"u, Nucl. Phys. Proc. Suppl. 74, 227-230 (1999);
Y.-H. Chen, H.-Y. Cheng, B. Tseng, K.-C. Yang,
  Phys. Rev. D60, 094014 (1999); H.-Y. Cheng and K.-C. Yang,
     Phys. Rev. D62, 054029 (2000).

 \bibitem{bbns}M. Beneke, G. Buchalla, M. Neubert, C.T. Sachrajda,
 Phys. Rev. Lett. 83, 1914 (1999); Nucl. Phys.
 B591, 313 (2000).

 \bibitem{Keum:2000} Y.-Y. Keum, H.-n. Li and A. I. Sanda,
 Phys. Lett. B504, 6 (2001); Phys. Rev. D63, 054008 (2001);
 C.-D. L\"u, K. Ukai and M.-Z. Yang,
 Phys. Rev. D63, 074009 (2001); C.-D. L\"u,   page  173-184,
 Proceedings of
     International Conference on Flavor Physics (ICFP 2001), World Scientific,
     2001,
  hep-ph/0110327.
 \bibitem{Chang:1997dw} C.-H. V. Chang and H.-n. Li, Phys. Rev. D55, 5577 (1997);
 T.-W. Yeh and H.-n. Li, Phys. Rev. D56, 1615 (1997).

 \bibitem{PQCD}
 H.-n. Li, Phys. Rev. D64, 014019 (2001);
 S. Mishima, Phys. Lett. B521, 252 (2001);
 E. Kou and A.I. Sanda, Phys. Lett. B525, 240 (2002);
 C.-H. Chen, Y.-Y. Keum, and H.-n. Li, Phys. Rev. D64, 112002 (2001);
 C.-D. L\"u and M.Z. Yang, Eur. Phys. J. C23, 275 (2002);
 A.I. Sanda and K. Ukai, Prog. Theor. Phys. 107, 421 (2002);
 C.-H. Chen, Y.-Y. Keum, and H.-n. Li, Phys. Rev. D66, 054013 (2002);
 M. Nagashima and H.-n. Li, hep-ph/0202127;
 Y.-Y. Keum, hep-ph/0209002; hep-ph/0209208(to appear in PRL); hep-ph/0210127;
 Y.-Y. Keum and A. I. Sanda, Phys. Rev. D67, 054009 (2003);
 C.D. L\"u, M.Z. Yang, hep-ph/0212373, to appear in Eur. Phys. J. C..

 \bibitem{L3} H.-n. Li, Phys. Rev. D66, 094010 (2002).

 \bibitem{soft} H.-n. Li and B. Tseng, Phys. Rev. D57, 443, (1998).
   \bibitem{lu}C.D. L\"u, Eur. Phys. J. C24, 121 (2002).
    \bibitem{lu2} C.-D. L\"u, K. Ukai, hep-ph/0210206, to appear at Eur. Phys. J. C; Y. Li,
    C.D. L\"u, hep-ph/0304288.

 \bibitem{h2h} T. Kurimoto, H.-n. Li, and A. I. Sanda, Phys. Rev. D67, 054028 (2003).
 \bibitem{Ball:1998je} P. Ball, JHEP, 09, 005, (1998); JHEP, 01, 010, (1999).

 \bibitem{Krokovny:2002pe} Belle Collaboration, P. Krokovny {\it et al.},
 Phys. Rev. Lett. 89, 231804 (2002).

 \bibitem{Aubert:2002eu} BaBar Collaboration, B. Aubert {\it et al.}, hep-ex/0207053.

   \bibitem{pdg} Review of Particle Physics, K. Hagiwara {\it et al.}, Phys. Rev. D66,
   010001 (2002).
 \bibitem{cl}  G.P. Lepage and S.J. Brodsky, Phys. Rev. D22, 2157
               (1980); J.D. Bjorken, Nucl. Phys. B (Proc. Suppl.) 11, 325
               (1989);
 C.-H. Chen and H.-n. Li, Phys. Rev. D63, 014003
 (2001).




\end{thebibliography}
\end{document}